\documentclass[prl,showpacs,twocolumn,floatfix, groupaddress,superscriptaddress,footinbib]{revtex4-1}
\usepackage{times,amsmath,amsfonts,amssymb,latexsym}
\usepackage{graphicx,epsf}
\usepackage{subfigure}
\newcommand{\be}{\begin{equation}}
\newcommand{\ee}{\end{equation}}
\newcommand{\ba}{\begin{eqnarray}}
\newcommand{\ea}{\end{eqnarray}}
\newcommand\tr{{\mbox{Tr\,}}}
\newcommand{\ignore}[1]{}

\newcommand{\ket}[1]{\left | {#1} \right \rangle }

\newcommand{\bra}[1]{\left \langle {#1} \right | }

\def\CC{{\rm\kern.24em \vrule width.04em height1.46ex depth-.07ex
    \kern-.30em C}}
\def\P{{\rm I\kern-.25em P}}
\def\RR{{\rm
         \vrule width.04em height1.58ex depth-.0ex
         \kern-.04em R}}

\def\bbbc{{\mathchoice {\setbox0=\hbox{$\displaystyle\rm C$}\hbox{\hbox
to0pt{\kern0.4\wd0\vrule height0.9\ht0\hss}\box0}}
{\setbox0=\hbox{$\textstyle\rm C$}\hbox{\hbox
to0pt{\kern0.4\wd0\vrule height0.9\ht0\hss}\box0}}
{\setbox0=\hbox{$\scriptstyle\rm C$}\hbox{\hbox
to0pt{\kern0.4\wd0\vrule height0.9\ht0\hss}\box0}}
{\setbox0=\hbox{$\scriptscriptstyle\rm C$}\hbox{\hbox
to0pt{\kern0.4\wd0\vrule height0.9\ht0\hss}\box0}}}}
\def\bbbz{{\mathchoice {\hbox{$\sf\textstyle Z\kern-0.4em Z$}}
{\hbox{$\sf\textstyle Z\kern-0.4em Z$}}
{\hbox{$\sf\scriptstyle Z\kern-0.3em Z$}}
{\hbox{$\sf\scriptscriptstyle Z\kern-0.2em Z$}}}}

\newcommand{\PRL}{Phys.~Rev.~Lett.}

\usepackage{hyperref}

\begin{document}

\title{Quantum entanglement in random physical states}

\author{Alioscia Hamma}

\affiliation{Perimeter
Institute for Theoretical Physics, 31 Caroline St. N, N2L 2Y5,
Waterloo ON, Canada}

\author{Siddhartha Santra}
\affiliation{Department of Physics and Astronomy and Center for Quantum Information Science and Technology,
University of Southern California, Los Angeles, California 90089-0484, USA}

\author{Paolo Zanardi}
\affiliation{Department of Physics and Astronomy and Center for Quantum Information Science and Technology,
University of Southern California, Los Angeles, California 90089-0484, USA}
\begin{abstract}
Most states in the Hilbert space are maximally entangled. This fact has proven useful to investigate -- among other things -- the foundations of statistical mechanics. Unfortunately, most states in the Hilbert space of a quantum many-body system are not physically accessible. We define physical ensembles of states acting on random factorized states by a circuit of length $k$ of random
and independent unitaries with local support. We study the typicality of entanglement by means of the purity of the reduced state. We find that for a time $k=O(1)$, the typical purity obeys the area law. Thus, the upper bounds for area law are actually saturated, on average, with a variance that goes to zero for large systems. Similarly, we prove that by means of local evolution a subsystem of linear dimensions $L$ is typically entangled with a volume law when the time scales with the size of the subsystem. Moreover, we show that for large values of $k$ the reduced state becomes very close to the completely mixed state. 
\end{abstract}

\maketitle

{\em Introduction.---} Entanglement is the defining characteristic of quantum mechanics: it is a key ingredient in quantum information processing\cite{ent-qip}, quantum many-body theory \cite{amico}, and the description of novel quantum phases of matter \cite{topentropy}. 
More recently, quantum entanglement has shed new light on the foundations of  statistical mechanics, and  the processes of equilibration and thermalization. 
The idea consists in the fact that even with unitary evolution, if the entanglement is large enough,  the expectation values of local observables are typically close to those of the thermal state \cite{qterm, therm, winter1}. It has been shown that such typicality is related to the volume law for the entanglement in random states \cite{winter1}. 
The problem with this approach is that random states are not physical because they are not accessible in nature. Indeed,  one needs a doubly exponential time in the system size to access all the states of the Hilbert space. For this reason, some authors have argued that the Hilbert space is an illusion \cite{verstraete}. 

Nevertheless, physical states do thermalize, as  has been shown in experiments with cold atoms, theoretical models and numerical simulations \cite{thermquench, rigol, cramer08}, or show typicality in the expectation value of local observables \cite{garnerone}. Does this mean that the mechanism for thermalization is not entanglement? There are several examples of physical relevance showing that when  evolution time scales with the size of the system, the state is entangled with a volume law \cite{cc, eisert}. Can thus we prove any statement about the typicality of such situations? 

In this Letter, we propose to answer the following question: how much are typical {\em physical} states entangled? We adopt the 2-Renyi entropy as a quantum entanglement quantifier \cite{renyi} as opposed to von Neumann Entanglement Entropy (EE). While from the technical point of view this choice allows a drastic simplification of the theoretical treatment, on physical grounds, we expect all the scaling results for typical physical states presented in this Letter to be fulfilled by EE as well any other sound quantum entanglement measure.

To this end, we define an ensemble $\mathcal E^{(k)}$ of physical states in this way:
pick a product state of a multipartite system, then act with $k$ independent random unitaries, each of them compatible with some locality structure, e.g., supported on edges of a graph. This mimics the continuous evolution generated over a time $k$ by a local (time dependent) Hamiltonian and are amenable to an elegant analytical treatment. Indeed, by applying the group theoretic techniques  of Ref. \cite{entpower},  we can compute the ensemble  average and variance of the $2-$Renyi entropy $S_2$ of a  subregion $A$ with boundary of size  $|\partial A|$. The result is that, typically, for $k=O(1)$ we obtain an area law that is an entanglement  $S_2=O(|\partial A|)$, while for $k$ scaling as the linear size $L$ of the system the average purity shows a volume law. Moreover, we show that fluctuations are small, and that there is measure concentration around the average value. As a final result, we show that for $k\rightarrow\infty$, the subsystem typically reaches the completely mixed state. 

Note that the upper bounds to entanglement laws that incorporate the locality of the interactions are known. Using the technique of the Lieb-Robinson bounds, one can prove that entanglement that can be produced in a subsystem $A$ by evolving for a time $t$ is upper bounded by a quantity scaling with $|\partial A|\times v t$. Here $v$ is limit speed for the interactions in the system \cite{lrb}. In other words, for $t=O(1)$ the area law for the entanglement is an upper bound, while the volume law is an upper bound for $t=O(L)$. These upper bounds have been proven very useful, e.g., in the context of simulability of quantum many-body systems or the understanding of topological order. However, these are just upper bounds and they deal with an extremal case. It could be the case that most Hamiltonians are  very weakly entangling. Therefore one wonders how typical area and volume laws are. Are these upper bounds  saturated on average and how strong are the fluctuations around the average? 
  The results of this Letter say that almost every local evolution entangles the subsystem with a scaling law $S_2=O(|\partial A|\times t)$. 

A second motivation is found in the context of unitary $t-$ designs \cite{harrow}.
Known results scale with some polynomial of the total number $n$ of degrees of freedom \cite{harrow, recent}. In this Letter, we focus on the statistics of observables of a reduced system, and the asymptotic results scale with the size of the reduced system.


{\em Statistical ensemble of physical states.---} 
 We start by defining the ensembles $\mathcal E^{(k)}$; henceforth we will refer to the elements of these ensembles as the  physical states.
Let $V$ be a set  of vertices endowed with a probability measure   $p: X \subset V  \mapsto p(X)\in[0,1]$ with $\sum_{X}p(X)=1$.
To each of the vertices $x\in V$ we associate a local $d-$dimensional Hilbert space $\mathcal H_x\simeq \mathbb C ^d$. The total Hilbert space is thus $\mathcal H_V \equiv \otimes_{x\in V} \mathcal H_x$, or the space of $n$ qudits. A completely factorized state in $\mathcal H_V$ has the form $\ket{\Phi} = \otimes_{x\in V}\ket{\phi_x}$; let $\omega =\ket{\Phi}\bra{\Phi}$ be its density matrix. 
The statistical ensembles  of quantum states $\mathcal E^{(k)}$ are constructed in the following way: We first draw a subset $X\subset V$ according to the measure $p$ and then we draw a unitary $U_X\in\mathcal U(\mathcal H_X)$ according to a chosen  measure  $d\mu (U|X)$. Then we define
$\mathcal E (p, d\mu) = \{ U_X\ket{\Phi} \}_{X,U_X,\Phi}.
$
This, ensemble can be generalized to the $k-$iterated ${\mathcal E}^{(k)}$ by considering unitaries of the form $U=\prod_{i=1}^k U_{X_{i}}$ where the $X_i$'s ($U_{X_i})$'s are drawn according the product
probability $\prod_{i=1}^k p(X_i)$ ($\prod_{i=1}^k d\mu(U_{X_i})):$ at each tick $i$ of the clock a new independent set $X_i$'s and a unitary $U_{X_i}$ are picked. In this ensemble, we can compute the statistical moments of any Hermitian operator. It turns out that by varying over the $U_X$, one can pick all the possible factorized $\Phi$'s so that the $\Phi$ dependence can be in fact dropped.

{\em Subsystem purity.---}
We now examine the typical entanglement in the statistical ensembles ${\mathcal E}^{(k)}.$
Let us consider a bipartition $V= A\cup B$ in the system: $\mathcal H_V =\mathcal H_A\otimes \mathcal H_B$, where $\mathcal H_J = \otimes_{x\in J}\mathcal H_x$ with obvious notation. We take a state $\rho\in \mathcal E_k$ and consider the reduced density matrix $\rho_A = \tr_B (\rho)$. In order to evaluate the entanglement of $\rho$ we  compute the purity $P^{}=\mathrm{Tr}_A(\rho_A^{2})$ and thus the $2-$Renyi entropy $S_2=-\log P$. To compute this trace  we use the  well-known fact that the trace over the square of every operator can be computed as the trace of two tensored copies of that operator times the swap operator. Indeed, defining 
$\rho^{\otimes 2}=U^{\otimes 2} \omega^{\otimes 2}{U}^{\dagger\otimes 2}$ and considering the order 2 shift operator (swap) on  $T_x: \mathcal H_x^{\otimes 2}\mapsto\mathcal H_x^{\otimes 2}$, we 
have 
$P^{}=\mathrm{Tr}_A(\rho_A^{\otimes 2}\tilde{T}_A)=\mathrm{Tr}[\rho^{\otimes 2}T_A]
$
where 
$
\tilde{T}_A=\otimes_{x\in A}T_x :(\mathcal{H}_A)^{\otimes 2}\mapsto(\mathcal{H}_A)^{\otimes 2}$
by $\ket{i_1,i_2}\mapsto\ket{i_{2},i_1}$
is the order $2$ shift operator in the $A$ space $\mathcal{H}_A=\otimes_{i\in A}\mathcal{H}_i$ and $T_A:(\mathcal H_A\otimes\mathcal H_B)^{\otimes 2}\rightarrow (\mathcal H_A\otimes\mathcal H_B)^{\otimes 2}$ is  given by $T_A=\tilde{T}_A\otimes\openone_B.$ 
We can now consider different concrete  ensembles. As a  basic model, let us consider the case in which there is a just as single edge: 
the system $\mathcal E_{edge}$ consists of two sites $A=\{i\}$ and $B=\{j\}$ connected by an edge $e$ so that the Hilbert space is $\mathcal H_e =  \mathcal H_A \otimes \mathcal H_B$ of dimensions $d_A=d_B=d$. The probability distribution is the trivial $p(e)=1$ and we pick the unitaries $U_e (\mathcal H_e)$ with the Haar measure: $d\mu(U_e) =d\mu_{Haar}$. Notice that in this case, the "locality" does not play any particular role. There is just one edge so the unitaries $U_e$ are the unitaries over the whole $\mathcal U(4)$. 
Following Ref.\cite{entpower}, one can exploit the  group theoretic structure of the ensemble $\mathcal E$  to compute average and statistical moments of operators. The average of an operator over a group action is indeed the weighted sum of projectors onto the IRreps of the representation of that group.  A direct calculation (see supplementary material) shows that $\overline{P^{}}^U= {d^3}/{d_+}=2{d}/{(d^2+1)}\equiv 2N_d.$ For very large $d$, we approximate the completely mixed state. Since the purity is a positive definite quantity that is going to zero in the thermodynamic limit, in the limit for the large system the fluctuations are also very small. Notice that this result reproduces what we know: a random state in the whole Hilbert space is typically very entangled. 


{\em Propagation of typical entanglement.---} The main goal of this paper is to explore what is the typicality of entanglement when there are some local conditions on how the ensembles of states are constructed. The local conditions are implemented by acting $k$ times with random local quantum circuits. We show that the loss of purity in a subsystem $A$ propagates, {\em in average}, within a length $\sim k$ within the bulk of $A$. In this way, we can show that, in average, the entanglement for $k=O(1)$ follows the area law, and for a generic $k$ it follows $S_2=O(|\partial A|\times t)$. Moreover, we show that the variance of the distribution of entanglements is very small: the average entanglement is typical. We can then conclude that the laws that determine upper bounds for the entanglement actually also determine the typical situation. In order to obtain this kind of propagation result, we will exploit a result on the algebra of the permutations $T_A$ ($A\subset V$) defined above. Let us start by defining 
the superoperator that averages over the unitaries $U_X\in \mathcal{U}(\mathcal{H}_X)$, that is, 
$\mathcal R_X (T_A) = \int d\mu(U|X) (U_X^\dagger)^{\otimes 2} T_A (U_X)^{\otimes 2}.
$
Notice that when $d\mu$ is the Haar measure, the $\mathcal R_X'$s are projection superoperators; in the rest of the letter we will focus on this case.
Then we can evaluate the average purity  as 
\be
\overline{P}= \langle \omega^{\otimes 2}, \mathcal R (T_A)\rangle
\label{ave-purity}
\ee
where $\mathcal R =\sum_{X\subset V} p(X) \mathcal R_X $ is a self-dual (Hermitian) superoperator. As a far as the purity calculations are concerned this superoperator completely characterizes the ensembles $\mathcal E^{(k)}$. Indeed, it is now easy to see that --in view of the statistical independence of each iteration-- the average purity for the $k-$iterated ensemble ${\mathcal E}^{(k)}$ is given by the expression (\ref{ave-purity}) with $\mathcal R $ replaced by ${\mathcal R}^k.$   In order to understand the spectral properties of $\mathcal R$, observe that:
 $\|\mathcal R\|\le \sum_{X\subset V} p(X) \|\mathcal R_X\| \le \sum_{X\subset V} p(X)=1.$  Since $\mathcal R (\openone)=\openone$ we then see that 
$\|\mathcal R\|=1$ whence the eigenvalues $\lambda_\alpha$of $\mathcal R$ are bounded in modulus by one and the highest one is $\lambda_1=1$. One can then write $ \bar{P}_k=
\sum_\alpha \lambda_\alpha^k c_\alpha;$  where 
$c_\alpha:=\langle \omega^{\otimes 2}, (T_A)_\alpha\rangle$ and  $(T_A)_\alpha$ denotes the projection of $T_A$
onto the eigenvalue $\lambda_\alpha$ eigenspace of $\mathcal R.$ For $k\to\infty$ this quantity goes to the limite value $c_1$ while the convergence rate is dictated  the second highest eigenvalue $\lambda_2$ of $\mathcal R$ \cite{hsz2}.

One of the key steps to obtain the results of this Letter is to realize that the $\mathcal R$'s superoperators can be regarded as maps
on the $2^{|V|}$-dimensional space spanned by the $T_X$'s ($ X\subset V$) into  itself (instead of maps of the $d^{4|V|}$-dimensional ${\mathcal L}({\mathcal H}_V^{\otimes 2})$ into itself).
For example, if   $X:=\{a,b\}$ i.e., an edge and $A$ is any subset of $V,$  a calculation similar to the above leads to
\ba\nonumber
\mathcal R_X(T_A)& = &  N_d (T_{A\backslash X}+T_{A\cup X})\; X\cap A \ne\emptyset \wedge X\cap B\ne\emptyset \\
\mathcal R_X(T_A)& = &  T_A \qquad \text{otherwise}
\label{algebra}
\ea
The edge $\mathcal R_X$'s have low-dimensional invariant subspaces of permutations, e.g., in a chain topology the span of the $T_A$'s associated with connected $A$'s is invariant.
This remark along with the fact that  $\langle \omega^{\otimes 2}, T_X\rangle=1$ for product states, allows for drastic simplifications in the evaluation of the average purity of $\mathcal E^{(k)}.$ The content of Eq.(\ref{algebra}) is that $\mathcal R_X$' has a non trivial action only if the edge $X$ straddles the boundary between the system $A$ and $B$. 

As an example, let us show how  the algebra (\ref{algebra}) simplifies the calculation of  the average purity for the single edge model. The subsistem $A$ is just one site and therefore  $A\backslash e=\emptyset$ and $A\cup e= \{A,B\}$. Moreover, $T_\emptyset=I$ and $\mathcal R_e (T_A) = N_d(I+T_AT_B)$. Finally we get
$\overline{P}= \langle \omega^{\otimes 2}, \mathcal R_e(T_A)\rangle = 2N_d$.
In the case of qubits, $d=2$ and $2N_2=.8$. It is also possible to compute the variance by generalizing the group averages to higher power of the density operator to obtain $\Delta P=.017$. A systematic treatment is to be found in \cite{hsz2}.


At this point, we consider a system with a notion of locality, so that it makes sense how  average entanglement  propagates. To this aim, we define the   $k$-random edge model $\mathcal E^{(k)}_{random}$ on a graph $G=(V,E)$ where $V$ is the set of the nodes and $E$ the set of the edges. 
We define a flat probability distribution on the edges of the graph $\Gamma$: $p(X) = 1/|E|$ if $X\in E$ and zero otherwise. Then we pick the unitaries on the edges with the Haar measure: $d\mu(U|X) =d\mu_{Haar}$.
We call $\partial A\subset E$ the subset of edges that have nonnull intersection with both $A$ and $B$. The probability of an edge to belong to $\partial A$ is thus $q= |\partial A|/|E|$. We are interested in the thermodynamic situations where $q\ll 1$. Using  Eq.(\ref{algebra}) we get
\be
\mathcal R (T_A)=\!\!\sum_{X\in E\backslash \partial A}p(X)T_A + \sum_{X\in \partial A}p(X)N_d [T_{A\cup X} + T_{A\backslash X}]
\label{random}
\ee
where $X\in E$ is an edge of the graph. One can see that  only the terms in Eq.(\ref{random}) that live across the boundary will decrease the purity of the subsystem. Moreover, the  support of $\mathcal R (T_A)$ is now on graphs with locally modified boundaries. 
For $k=1$ is then easy to find the average purity:
 $\overline{P} = \sum_{X\in E\backslash\partial A} p(X)\langle \omega^{\otimes 2},T_A\rangle +\sum_{X\in\partial A} N_d\langle \omega^{\otimes 2}, T_{A\cup X} + T_{A\backslash X} \rangle= 1-(1-2N_d)q.$ 
 From Eq.(\ref{random}) we see that every application of $\mathcal R_X$ transforms the subset $A$ into a superposition of $A\cup X$ and $A\backslash X$ so that at any successive iteration the boundary of the new subset changes and its boundary length may change. 
The iteration for $k$ scaling with the linear size of the system gives the results (see supplementary material)  
$\overline{P_k}\simeq (1-q(1-2N_d))^k$. Therefore the average $2-$Renyi entropy of the ensemble is $\bar{S}_2(k) \equiv -\overline{\log P_k}\ge  -\log\overline{P}_k\simeq -k\log(1-q(1-2N_d))\simeq (1-2N_d)qk$. In terms of the average entangling power  \cite{entpower}, one gets 
$\bar{S}_2(k)\ge k {|\partial A|}/{|E|}\,\overline{e_p(U)}^U.
$
In other words, the average $2-$Renyi entropy for the random edge model of the $k$th iteration is lower bounded by $k$ times the fraction of vertices in the boundary of the region $A$ times the average entangling power of an edge unitary, showing a linear increase of entropy in time, or, in other words, {\em the entanglement is propagating into the bulk of $A$}. 
Moreover, one can compute variances of $\overline{P}$ and show that $\sqrt{\Delta S_2}/S_2 \sim 1/\sqrt{|\partial A}|$ \cite{hsz2}.  This in turn implies measure concentration (typicality) in the thermodynamic limit $|\partial A|\rightarrow\infty$. 
So a random circuit model of this type can reproduce the Haar measure for the statistics of observables on the reduces system, as $k$ scales with the subsystem size.


\begin{figure}
  \centering
  \includegraphics[scale=.3]{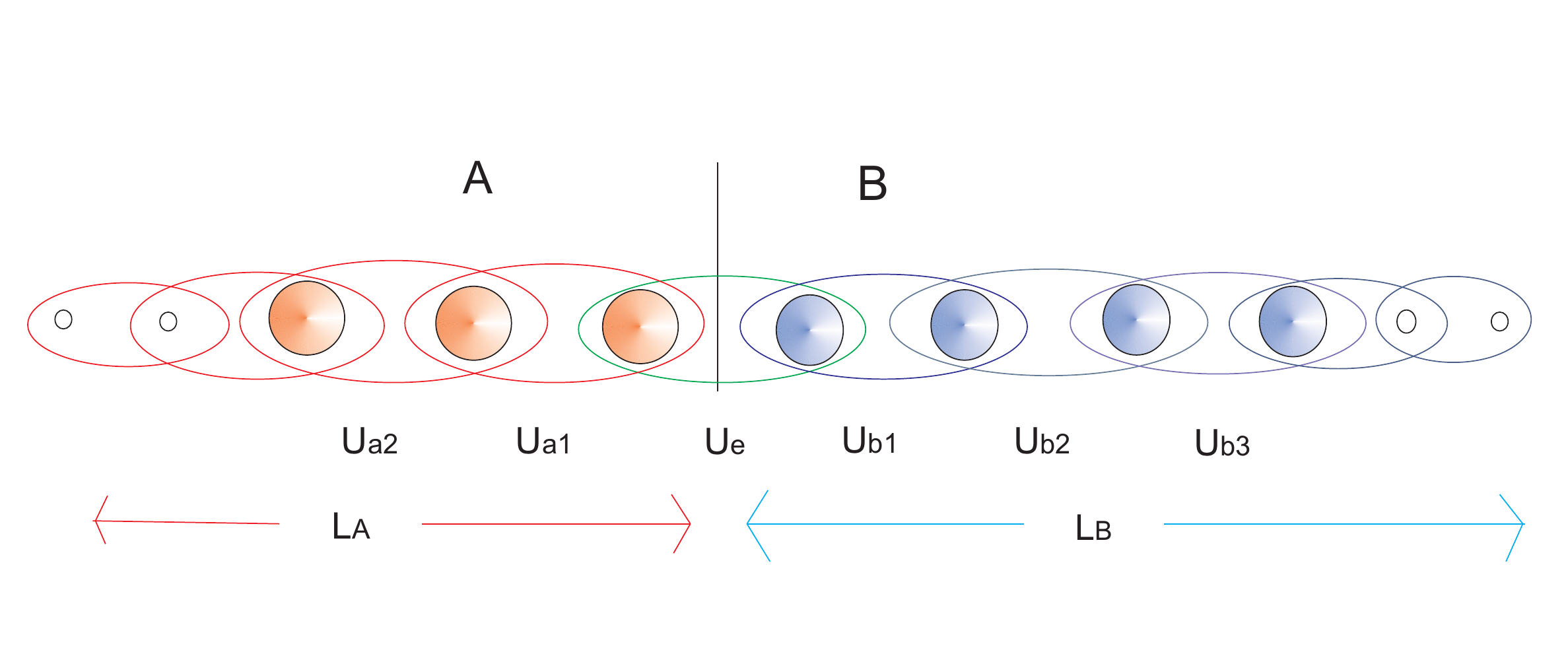}
  \caption{A bipartite $(A,B)$ spin chain of length $L=L_A+L_B$ with nearest-neighbor qubits interacting via $2-$qubit gates (ellypses). The edge $e$ is the one that straddles the two partitions. The gates are numbered by the subscript $ai,bi$ where $i$ is the distance from the boundary.}
  \label{linchain}
\end{figure} 



{\em The linear chain.---} We now move to the case corresponding to a time dependent Hamiltonian that is the sum of local terms.
In this model, the unitaries act on all the edges of the graph $\Gamma.$
The probability distribution is thus $p(X)=1$ for $X=V$ and zero otherwise. For the sake of simplicity
in the following, we will consider the case of the graph $\Gamma$ being a bipartite chain of length $L=L_A+L_B$. 
Extensions to higher dimensional geometries will be presented in \cite{hsz2}).  We will label by $U_e$ the unitary acting on the edge straddling the $(A,B)$ bipartition, while we will use the labels $a_i, b_i$ for the unitaries that act in the bulk of $A,B$ respectively (see Fig.\ref{linchain}).  We label the sites of the chain as $L_A,..,1_A,1_B,...L_B$. Since the unitary is a product of all the edge unitaries, we need to specify in which order they act. In the following, the unitary $U_\sigma$ will always denote the product over all the edges in $E$ with the order given by the permutation $\sigma$, so $U_\sigma$ is the ordered product over local two-qudit unitaries. This corresponds to the (time ordered) infinitesimal evolution with a local Hamiltonian, where $\sigma$ gives the time ordering:
$U_\sigma = U_{\sigma(e_1)}  \ldots U_{\sigma(e_{|E|})}$. At this point we 
construct the set $\mathcal E_{(\sigma)}^{(k)} (\Gamma)=\{U_\sigma\ket{\Phi}\}_{U}$ with measure $d\mu(U) = \delta (U-U_\sigma)\prod_{e\in E} d\mu_{Haar}(U_e)$.  
This ensemble approximates all the states that can be evolved from a factorized state with a local Hamiltonian acting for an infinitesimal amount of time. By $k$ iteration, we
 obtain the time evolution for a finite time $k:$
$\mathcal E^{(k)} _{chain}(\Gamma) =\{ \prod_{i=1}^k U_{\sigma_i} \ket{\Phi}\}_{U, \sigma}.
$
Here,  we consider all the possible ordered sequences of unitaries by taking, a each time step, a  permutation $\sigma$ of the edges uniformly at random. 
This ensemble approximates all the states that can be reached in time $k$ by the evolutions originated by all the possible  time-dependent Hamiltonians on a graph. The ensemble $\mathcal E^{(k)}$ thus only depends on the number of iterations (the "time") $k$ and the graph $\Gamma$. The loss of purity due to the action of the unitaries depends on their order. Thus, 
in order to find an upper bound to the average purity, we consider the ordering that gives the minimum loss of purity. As $k$ increases, nodes at distance $k$ from the boundary participate in the averaging calculation. A lengthy calculation shows that the purity gets a factor $N_d$ for every node participating in the average  and we find (see supplementary material)
 $\bar{P}_{k}=\sum_{m=0}^{m=k-1}2 \binom{k+m-1}{m}N_d^{(k+m)}$. Summing the series for  $\bar{P}_{k}$ for large values of $k$ one finds 
 $\bar{P}_k\simeq2\left[{N_d}/{(1 - N_d)}\right]^k$.
 Recall that this equation, in view of the choice of the $\sigma$ corresponding to the $U_\sigma$ with the least entangling power, is an upper bound for the purity in $\mathcal E^{(k)}_{chain}.$
The exponential decay of the purity in $k$ is due to the fact that all the qudits at distance $k$ from the edge are getting mixed. 
Since average $2-$Renyi entropy is $\bar{S}_2\ge=-\log \bar{P}_k,$ we have the lower bound 
\be
\overline{S_2}\ge k\log\left(\frac{1-N_d}{N_d}\right)-\log2\simeq k\log d -\log 2
\label{S2}
\ee
(Last approximation holds for large $d$). Eq. (\ref{S2}) for  $k=O(L_A)$ implies a) a {\em volume law} for the entanglement scaling  b)  typicality: a nearly minimal value of the average of purity (in view of the Markov inequality), forces also the fluctuations around this average to be  small.
For $k>|L_A|$ one has no longer a linear increase of entanglement with time  but observes a saturation.  This type of behavior has been found in examples
of entanglement dynamics after a quench using CFT  techniques \cite{quench2, cc,recent}.   

To study the  limit of average purity for  $k\rightarrow\infty$  we first notice that  the chain superoperator is a ($\sigma$-ordered) product of (non-commuting) projections 
$\mathcal R_{chain}=\mathcal  R_{\sigma(|E|)}\cdots\mathcal  R_{\sigma(1)}$ .
This implies $\|\mathcal R_{chain}\|\le \prod_{e\in E} \|\mathcal R_e\|\le 1;$ again this means that all the eigenvalues  of $\mathcal R_{chain}$ are smaller in modulus than $1$
and therefore asymptotically just fixed points e.g.,$\openone, T_V,$ contribution to $T_A$ survives.
If now one {\em assumes} that the symmetric combination $\openone +T_V$ is the only relevant fixed point one finds $\bar{P}_{k\to\infty}={(d^{2L-L_A} + d^{L+L_A})}/{d^L(d^{L}+1)}$. We have checked this result by numerical simulations  \cite{hsz2} for the least and most entangling $\sigma$'s but we conjecture it to hold true for all orderings
and besides the one-dimensional chain scenario.
For large $|V|=L$ one has  $\bar{P}_{k\to\infty }\simeq d^{-L_A} +d ^{L_A-L}$ that in turn for $L_A\le L/2$
shows that the asymptotic purity differs from that of the totally mixed state $\openone_A/d^{L_A}$ for terms of order $d^{-L_B}.$
Finally, if $L_B\gg 1,$ this implies that the vast majority of the states in ${\mathcal E}^{(k)},$ once reduced to $A,$ are close in $L_1$-norm to the maximally mixed state.


{\em Conclusions.---} We investigated the typical entanglement in physical states. 
To this end, we defined statistical ensembles of physical states by considering product states on a multipartite system and evolving them with $k$ independent stochastic local gates. Ensemble averages can be computed by introducing suitable superoperators and using group-theoretic tools as in \cite{entpower}.
We would like also to stress that althoughin this Letter we  used purity to quantify the entanglement, this method extends in a straightforward way to general $\alpha$-Renyi entropy by natural modifications of superoperator $R$ and permutation $T_A$ in Eq.(\ref{ave-purity}) \cite{hsz2}. Assuming Ênow that one is allowed to perform an analytic Êcontinuation in the limit $\alpha\to1^+$, our results apply also for the von Neumann Entanglement Entropy.
States that are obtained by local evolution for a constant O(1)  time have a  typical entanglement given by the area law. While the area law was known to hold as an upper bound, we have shown that it is indeed {\em typical}. On the other hand, states that are obtained by evolution for a time scaling with the size of the system are shown to almost always obey the volume law for entanglement like typical  (Haar) random states do.
At this point, we may speculate as to whether this result implies local thermalization for physical states.

{\em Acknowledgments.---} We thank M.P.~ M\"uller for useful discussions. Research at Perimeter Institute for Theoretical
Physics is supported in part by the Government of Canada through NSERC and
by the Province of Ontario through MRI. PZ acknowledges support from NSF grants PHY-803304 and  PHY-0969969.
This research is partially supported by the ARO   MURI grant W911NF-11-1-0268.

 \section{Supplementary Material}
 
\subsection{Single edge}\label{sem}
Let us show the detailed calculation of the average purity for the single edge model. In this case, the IRreps are carried by the totally symmetric ($\mathcal{H}^2_+$) and totally antisymmetric ($\mathcal{H}^2_-$) subspaces of $\mathcal{H}_e^{\otimes 2}$. 
The average is given by
$
\overline{P}^U= \int dU \tr[\rho^{\otimes 2}T_A]= \tr[\omega^{\otimes 2} \int dU ({U}^\dagger)^{\otimes 2} T_A U^{\otimes 2}]
$.
After the integration we get
$
\overline{P^{}}^U=d_+^{-1}{\mathrm{Tr}(\Pi_+~T)}\mathrm{Tr}(\omega^2 \Pi_+)=d_+^{-1}{\mathrm{Tr}(\Pi_+~T)}
$
since $\omega^{\otimes 2}$ is supported only in the totally symmetric subspace $\Pi_+$ whose dimension is $d_+={d^2(d^2+1)}/{2}$  and $\mathrm{Tr}[\omega^{\otimes 2} \Pi_+]=1$. 
The projector onto the totally symmetric space has the form
$\Pi_+={(\openone+T_i\,T_j)}/{2}\implies 1/2\mathrm{Tr}({(\mathbf{1}_{(i,j)^{\otimes 2}}+T_i\,T_j)}T_i)=d^3$
and we finally get
$\overline{P^{}}^U= {d^3}/{d_+}=2{d}/{(d^2+1)}\equiv 2N_d.
$
In  \cite{entpower2} it was defined the average entangling power $\overline{e_p(U)}^U$ as the average entanglement one attains from a factorized bipartite state by averaging over the unitaries in the whole space with the Haar measure. With this definition, 
$\overline{e_p(U)}^U:=1-\overline{P^{}}^U=1-2N_d=(d-1)^2/(d^2+1)$[see Eq. (5) in\cite{entpower2}].

\subsection{Iteration to $k$ in the Random Edge Model}
To understand the structure of  $\mathcal R^k (T_A)$ it is instructive to consider explicitly the $k=2$ case. 
 Iterating (3) and computing the purity using (1)  \cite{note}  we find $\overline{P} \simeq (1-q)^2 +(1-q)2q N_d+(2qN_d)^2=(1-q(1-2N_d))^2$ where we have used $|\partial (A\cup e)|\simeq |\partial (A\backslash e)|\simeq |\partial A|$. This is true e.g. if every vertex has degree $o(|\partial A|)$. The calculation easily extends to the case $k=o(|\partial A|)$ and thus one finds $\overline{P_k}\simeq (1-q(1-2N_d))^k$.
 
 \subsection{Iteration to $k$ for the linear chain}
 Let us first show what is the sequence that lower bounds the amount of average entanglement produced. We can see that, for $k=1$, acting in $A$ and $B$ after having acted on the edge $e$ does not change the purity, so a sequence of the type $U_A U_B U_e$ results in a minimal loss of purity. Moreover, the order of the unitaries inside $A$ and $B$ also counts. From the iteration of the algebra Eq.(2) we can see that if we pick the ordering in which we first act near the boundary and proceed towards the outer parts of the chain: $U_A U_B U_e$ where
$U_A=U_{a_{L_A}}U_{a_{L_A-1}}....U_{a_2}U_{a_1}$  and $U_B=U_{b_{L_B}}U_{b_{L_B-1}}....U_{b_2}U_{b_1}$ we will get the lowest possible powers of $N_d$ and correspondingly the least decrease of purity. As $k$ increases, the difference between different orderings is attenuated and for very large values of $k$ it can also be neglected. We will anyway always consider the worst case scenario of ordering $\sigma$ which corresponds to the minimal decrease of purity.

Now we want to show how the algebra Eq.(2) propagates the average entanglement in the linear chain. The action of the superoperator $\mathcal R$ in the linear chain model is more complicated because now $X$ is not just the support of one unitary, but it contains the ordered product of all the edges. In particular, notice that now $\mathcal R$ is not hermitean. Using Eq.(2) multiple times  we find: for $k=1$,  $\mathcal R(T_A)=N_d(T_{A-1}+T_{A+1})$. Where we used the notation $A+r=A\cup \{1_B,...,r_B\}$ and $A-r=A\backslash \{1_A,..,r_A\}$. At the second iteration $k=2$, we get $\mathcal R^2(T_A)=N_d^2T_{A-2}+2N_d^3T_{A-1}+2N_d^3T_{A+1}+N_d^2T_{A+2}$. We can see that nodes at distance $2$ from the boundary enter the expression. Each $T$ in the expression for $\mathcal R^k(T_A)$ gives a $1$ when we take the scalar product with $\omega^{\otimes 2}$. So we find $\overline{P}_{k=2}=2N_d^2+4N_d^3$. It is important to understand how the interactions propagate with $k$. A somewhat lengthy calculation shows that as $k$ increases, nodes at distance $k$ from the edge participate to the averaging procedure and for every node that participates we pick a power for the base $N_d$. For $k<L_A$ calculation gives  $\bar{P}_{k}=\sum_{m=0}^{m=k-1}2 \binom{k+m-1}{m}N_d^{(k+m)}$.


\begin{thebibliography}{30}
\bibitem{note} {Notice that $\langle \omega^{\otimes 2}, T_X\rangle=1, \forall X\subset V$ always for product states, which shows why the initial choice of the product state $\Phi$ is irrelevant.}
\end{thebibliography}
\end{document}